\documentclass[journal]{IEEEtran}
\usepackage{multirow}
\usepackage{footnote}
\usepackage{graphicx}
\usepackage{cite}
\usepackage{amsmath,amssymb,amsfonts}
\usepackage{multirow}
\usepackage{hyperref}
\usepackage{subcaption}
\usepackage{color}
\usepackage{soul}
\usepackage{textcomp}
\usepackage{array}
\usepackage{float}
\usepackage{stfloats}

\def\xx{\mathbf{x}}
\def\ww{\mathbf{w}}
\def\hh{\mathbf{h}}
\def\zz{\mathbf{z}}
\def\rr{\mathbf{r}}
\def\bb{\mathbf{b}}
\def\dt{\Delta{t}}
\def\WW{\mathbf{W}}
\def\UU{\mathbf{U}}
\def\RRR{\mathbb{R}}
\def\NNN{\mathbb{N}}

\def\yy{y_\tau}

\begin{document}

\title{Detecting Video Game Player Burnout  \\with the Use of Sensor Data and Machine Learning}

\author{
	\IEEEauthorblockN{
		\textsuperscript{1}Anton Smerdov,
		\textsuperscript{1}Andrey Somov,
		\textsuperscript{1}Evgeny Burnaev,
		\textsuperscript{2}Bo Zhou,
		\textsuperscript{2}Paul Lukowicz
		}
	
	\IEEEauthorblockA{
	\textsuperscript{1}Skolkovo Institute of Science and Technology, CDISE, Moscow, Russia}
	
	
	\IEEEauthorblockA{
	\textsuperscript{2}German Research Centre for Artificial Intelligence, Kaiserslautern, Germany}
	}


\markboth{IEEE Internet of Things Journal,~Vol.~x, No.~x, Month~2021}{Smerdov \MakeLowercase{\textit{et al.}}: Detecting Video Game Player Burnout}

\maketitle

\begin{abstract}
  Current research in eSports lacks the tools for proper game practising and performance analytics. The majority of prior work relied only on in-game data for advising the players on how to perform better. However, in-game mechanics and trends are frequently changed by new patches limiting the lifespan of the models trained exclusively on the in-game logs. In this article, we propose the methods based on the sensor data analysis for predicting whether a player will win the future encounter. 
  The sensor data were collected from 10 participants in 22 matches in League of Legends video game. We have trained machine learning models including Transformer and Gated Recurrent Unit to predict whether the player wins the encounter taking place after some fixed time in the future. For 10 seconds forecasting horizon Transformer neural network architecture achieves ROC AUC score 0.706. This model is further developed into the detector capable of predicting that a player will lose the encounter occurring in 10 seconds
 in 88.3\% of cases with 73.5\% accuracy.
  This might be used as a players' burnout or fatigue detector, advising players to retreat. We have also investigated which physiological features affect the chance to win or lose the next in-game encounter.
  
\end{abstract}

\begin{IEEEkeywords}
eSports, dataset, machine learning, psychophysiological assessment, sensing, video games
\end{IEEEkeywords}

\IEEEpeerreviewmaketitle


\section{Introduction}

In the last several years, eSports has progressed from casual entertainment to a fully-recognizable competitive discipline. In 2019 its global audience achieved 443 million and with total revenue of 958 million dollars \cite{newzoo_2020}.
Prize pools for top-tier eSports tournaments achieve tens of millions of US dollars, thus making competition between professional teams~\cite{esports-2020} very high. This creates a demand for tools aimed at training and analytics in eSports.

However, despite the recognition as a sport in many countries, eSports is still in the infancy period. There are no many
training methodologies or widely accepted methods for analytics.
From the players' point of view, it is unclear how to improve their skills in a particular game except for regular training,
watching how professional players perform, and communicating with the eSports community. These methods are pretty straightforward, and there is a lack of tools providing feedback about the player performance and instructing how to play better.

Current methods usually rely on the analysis of in-game data obtained from game logs, such as kills/deaths/assists events, abilities learned, players' positioning, and other in-game statistics. This information can be helpful in player skill assessment, performance evaluation, match outcome prediction, and post-match analysis. 
However, 
current methods usually rely on the analysis of in-game data only and do not consider data from the physical world. Physiological data may provide new insights into players' behavior. Sometimes these data may be more suitable for the eSports domain since models trained on in-game data only quickly become obsolete when a new patch is released.
Sensor data can supplement logs obtained from in-game data
to provide additional information for predictive models and potentially improve their performance.



A major flaw of the models trained on in-game data only is their short lifespan. These models capture tight dependencies between in-game parameters and targets, but when rules or game mechanics change, previous connections might not be preserved. Since patches and updates in eSports are systematic, models trained on in-game data should be constantly updated, which is not convenient in practice. In contrast, the  models trained on real-world sensor data do not depend on the game updates much; thus, they are more robust, long-lasting, and production-friendly. 

Although the usage of sensor data in eSports has been previously explored, prior research usually focuses on the data from one sensor only~\cite{eye_tracking_patterns_1}. It significantly 
limits the insights explored since the multi-sensor dependencies are not captured. Also, using a single sensor it is not possible to figure out which sensors are more important.

In this paper, we present an analysis of sensor data collected from 10 participants playing League of Legends. 
Data collected include chair movements, hand, and head movements, pulse, saccades, muscle activity, galvanic skin response, mouse and keyboard activity, facial skin temperature, Electroencephalogram (EEG), environmental temperature and relative humidity, $CO_2$ level, and pulse oximeter data. 
We utilize this data 
to predict outcomes of encounters extracted from in-game logs using machine learning algorithms. 
The best models achieve a 0.706 ROC AUC (AUC: Area Under the ROC Curve, ROC: Receiver Operating Characteristic Curve) score in forecasting whether a player will win the encounter taking place 10 seconds later in the game. It also achieves precision of 73.5\% with a recall of 88.3\% in detecting that a player will lose the next encounter. In other words, the model can detect if the player will lose the fight occurring in 10 seconds in 88.3\% of cases with 73.5\% accuracy.



A contribution of this paper is the collection of eSports related physiological data, their processing, and consequent analysis for both professional and amateur players. We describe the whole pipeline from the data collection to the interpretation of predictions made by machine learning and deep learning models. In particular, we propose a sensor network architecture for data collection in the eSports domain, a methodology for the experiment organization, techniques to extract relevant features from the data, automatic methods for encounters extraction and annotation from in-game logs, and models for encounter outcome prediction. 

The novelty
is (i) experimentation with the professional eSports athletes and amateur players specializing in League of Legends discipline, (ii) data collection from the entire team, and (iii) prediction of player performance relying solely on the sensor data.




This paper is organized as follows: in Section~\ref{related_work}  we overview current methods utilizing sensor data for eSports analytics; in Section~\ref{sensing_platform} we present a sensing platform for data collection and methodology;
in Section~\ref{data_analysis} we describe data preprocessing and a method for automatic encounter annotation; in Section~\ref{machine_learning} we overview machine learning algorithms and their results; in Section~\ref{conclusion} we discuss the results and provide concluding remarks.

\section{Related Work} \label{related_work}

Current research in eSports is mainly formed by a wide group of studies concentrated on in-game data analysis and small scope of work focused on the analysis of psychophysiological data in eSports. In this section, we cover both types of studies with a primary focus on research involving sensor data.

\subsection{Sensor Data in eSports Research}

Despite the growing popularity of eSports research involving psychophysiological data in the last several years, there is no much prior work in this field. The main challenges are the
difficulty
and expensiveness of data collection and annotation, lack of professional players involved in the research, and sensor data processing.

Authors in \cite{esports_visual_fixations} propose to use data from eye tracker to describe traits of professional players in terms of visual fixations, or saccades. They argue that pros have more variable fixation durations than amateur players. The authors also identified that the difference between ambient and focal fixations are more expressed for high-skilled players.
A similar idea is explored in \cite{gaze_reactions}, where authors managed to classify players into professionals and amateurs using reaction time extracted from the eye tracker data. They estimate the total reaction time as a sum of saccadic latency, the time between the saccade and fixation, and the time for aiming and shooting.
Research in \cite{eye_tracking_patterns_1, eye_tracking_patterns_2} also supports that Pro-players eye-movement patterns are more diverse and swifter.

Another idea is to use the EEG data to figure out how the  players react to game events. In \cite{esports_eeg_0} authors demonstrated that negative game events decrease EEG alpha rhythms power spectra frequencies, while positive game events increase it. The authors also showed the opposite relation to the theta rhythm.
Work by Meneses-Claudio et al. \cite{esports_eeg_1} identified that a professional player has fewer variations in the EEG signal than an amateur player.

Activity on a chair can also be used to separate experienced and amateur players. Work by Smerdov et al. \cite{smart_chair_wf_iot} 
showed that features obtained from accelerometer and gyroscope integrated into a game chair provide a valuable signal about player skill in Counter Strike: Global Offensive (CS:GO). 
This work is extended in \cite{smart_chair_iop}, where authors additionally investigated how players react to game events in terms of chair movements. It turned out that professionals do not react to game events as much as newbie players.
Also, the authors showed that the Pro-players moves in a chair less frequently, but their movements are swifter. The state-of-the-art wireless technologies~\cite{nb-iot} ensure flexibility for the experimental testbed.

An abundant and natural source of data for eSports is mouse and keyboard activity. 
Authors in \cite{esports_skill_prediction} investigated the connection between players’ skill in CS:GO and data from mouse, keyboard, and eye tracker. They identified
the frequency and duration of keyboard strokes as the most important features.
In \cite{esports_gsr} authors presented a dataset with mouse, keyboard, and galvanic skin response data for League of Legends players but did not draw any conclusions. Kaminsky et al. \cite{identification_by_mouse} showed that mouse activity might be used for player re-identification.


\subsection{In-game Data in eSports Research}

In-game activity is a common source of data for eSports research. It is utilized in many research directions, including skill assessment, match outcome prediction, toxic behavior detection, etc.

A straightforward usage of in-game data is match outcome prediction. Several studies showed that the information about hero draft could be used to predict the match outcome in the Multiplayer Online Battle Arena (MOBA) genre \cite{lol_draft_prediction_0,dota_draft_prediction_1, dota_draft_prediction_2,dota_draft_prediction_0}. Such models have a direct practical application as draft recommendation systems.

Another possibility to predict the match outcome is to use in-game statistics like kills, deaths, gold, abilities learned, etc. This approach showed that, to some extent, match outcome could be predicted in the MOBA genre \cite{lol_live_prediction_0, lol_live_prediction_1, dota_live_prediction_0, dota_live_prediction_1} and first-person shooter (FPS) genre \cite{fps_prediction_0}. More complex models also try to interpret model results \cite{lol_prediction_interpretation, lol_logic_mining}, or calibrate confidence of a prediction \cite{confidence_prediction_lol}.

A popular subfield of eSports research is toxic behavior detection. 
Studies in this area cover automatic toxicity detection with natural language processing techniques \cite{toxicity_detection}, investigation of how participants interact with toxic players \cite{toxic_behavior}.


\subsection{Sensor Data in Skill Assessment}

Sensor data is already used for skill assessment in other domains.

A popular domain for skill assessment by sensor data is surgery. Surgeon expertise is shown to be connected with data from eye tracker \cite{surgery_skill_sensors_2_eye_tracking}, pressure sensors \cite{surgery_skill_sensors_3_pressure_sensors}, and Inertial Measurement Units (IMUs) \cite{surgery_skill_sensors_0, surgery_skill_sensors_1}.

Physiological data is a natural source of information in sports.
Players' movements~\cite{barshan-2020} captured by accelerometer and gyroscope located on different body parts
might help to estimate player skill in tennis \cite{tennis_skill_0, tennis_skill_1}. Similar techniques utilizing data from IMU have been employed for skill estimation in soccer \cite{soccer_detection_classification}, golf \cite{sport_activities_skill_sensors_1}, climbing \cite{climb_skill_sensors}, volleyball \cite{volleyvall_skill_sensors}, and gym exercising \cite{mobile_exercise_assessment_0, mobile_exercise_assessment_1}.

Sensor data have also been used for estimating a skill level in other activities such as dancing \cite{dance_skill}, clay kneading \cite{worker_skills_clay_kneading}, and mold polishing \cite{worker_skills_mold_polishing}.

\section{Data Collection} \label{sensing_platform}

\subsection{Sensing Platform}


\begin{figure*}[!t]
    \centering
	\centerline{\includegraphics[width=\linewidth]{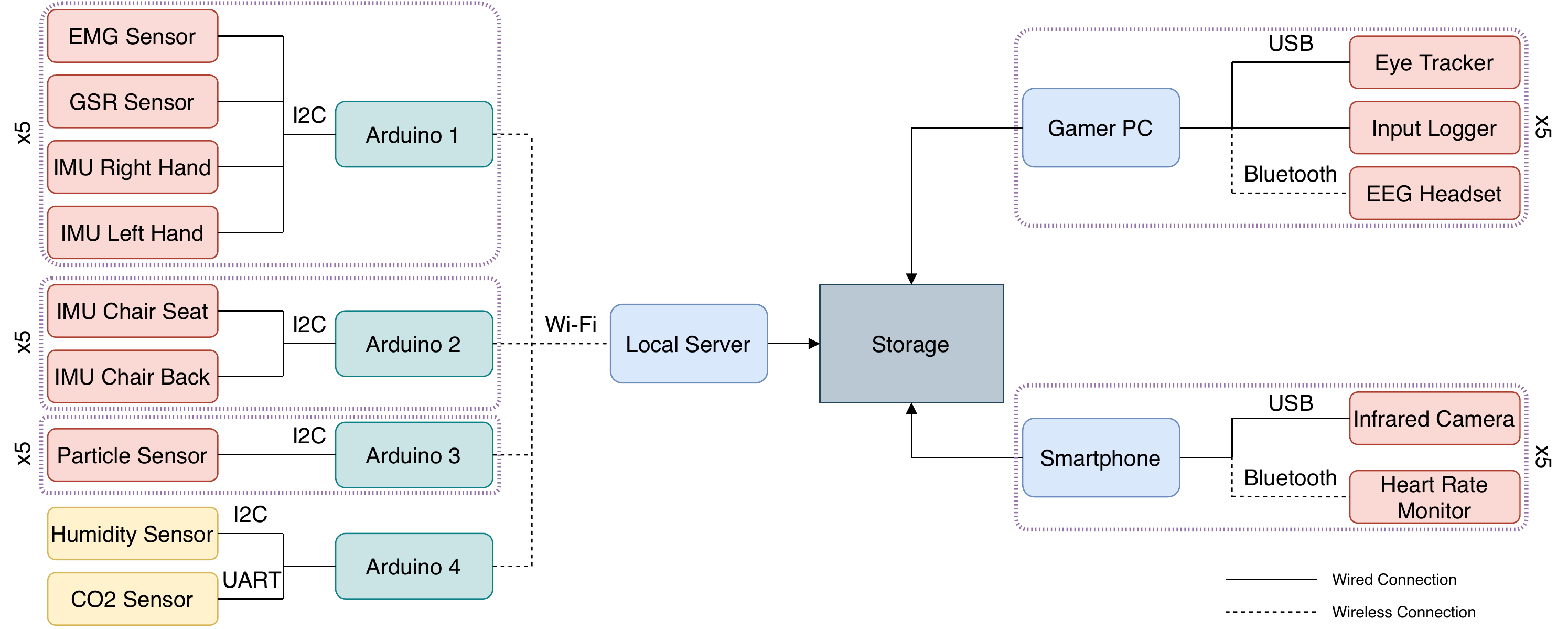}}
	\caption{Sensor network architecture.}
	\label{sens_arch}
\end{figure*}


\begin{table*}[!t]
\centering
\caption{Sampling rates for sensor data collected.}
\begin{tabular}{c|c|c|c|c}
\hline
Sensor name & Sensor model & Primary device     & Interface & Sampling rate, Hz   \\\hline\hline
EMG sensor, right hand & \multirow{2}{*}{Grove  EMG  Sensor v1.1\footnotemark} &  \multirow{5}{*}{Arduino 1} & \multirow{3}{*}{Analog}  & \multirow{5}{*}{36} \\\cline{1-1}
EMG sensor, left hand           &            &  & &                   \\\cline{1-1}\cline{2-2}
GSR sensor & Grove GSR Sensor v1.2\footnotemark &           &            &                     \\\cline{1-2}\cline{4-4}
IMU, right hand & \multirow{4}{*}{Bosch BNO055\footnotemark} & & \multirow{5}{*}{I2C}  & \\\cline{1-1}
IMU, left hand &   & & & \\\cline{1-1}\cline{3-3}\cline{5-5}
IMU, chair seat & & \multirow{2}{*}{Arduino 2} & & \multirow{2}{*}{65} \\
IMU, chair back & &          &  &  \\\cline{1-3}\cline{5-5}
Pulse-oximeter sensor & Maxim MAX30102\footnotemark     & Arduino 3 &   & 16-25\\\hline
CO$_2$ sensor & MH-Z19B\footnotemark & \multirow{2}{*}{Arduino 4} & UART  & \multirow{2}{*}{1}    \\\cline{1-2}\cline{4-4}
Environmental sensor & Bosch BME280\footnotemark        &  & I2C       &              \\\hline
Eye tracker & Tobii 4C  & Gamers' PC & USB & 90                  \\\hline
EEG Headset & Emotiv Insight\footnotemark          & PC        & Bluetooth & 8                   \\\hline
Thermal camera & Flir One \footnotemark & \multirow{2}{*}{Smartphone} & USB & 0.2 \\\cline{1-2}\cline{4-5}
Heart rate monitor & Polar OH1\footnotemark   & &  Bluetooth       & 1 \\
\hline
\end{tabular}
\label{sampling_rates}
\end{table*}
\setcounter{footnote}{0}
\stepcounter{footnote}\footnotetext{\url{https://wiki.seeedstudio.com/Grove-EMG_Detector/}}
\stepcounter{footnote}\footnotetext{\url{https://wiki.seeedstudio.com/Grove-GSR_Sensor/}}
\stepcounter{footnote}\footnotetext{\url{https://www.bosch-sensortec.com/products/smart-sensors/bno055.html}}
\stepcounter{footnote}\footnotetext{\url{https://www.maximintegrated.com/en/products/interface/sensor-interface/MAX30102.html}}
\stepcounter{footnote}\footnotetext{\url{https://www.winsen-sensor.com/d/files/infrared-gas-sensor/mh-z19b-co2-ver1_0.pdf}}
\stepcounter{footnote}\footnotetext{\url{https://www.bosch-sensortec.com/products/environmental-sensors/humidity-sensors-bme280/}}
\stepcounter{footnote}\footnotetext{\url{https://www.emotiv.com/insight/}}
\stepcounter{footnote}\footnotetext{\url{https://www.flir.eu/flir-one/}}
\stepcounter{footnote}\footnotetext{\url{https://www.polar.com/en/products/accessories/oh1-optical-heart-rate-sensor}}

To ensure proper data collection in an active gaming scenario, we developed a wearable sensing system comprising of several sensors.
The sensor network~\cite{wsn-iot-2020} architecture for data collection is shown in Fig.~\ref{sens_arch}.
Specific sensor models, connection interfaces, and sampling rates are shown in Table~\ref{sampling_rates}.

Each sensing device was controlled by a primary device, such as a PC or Arduino. 
The first group of sensors consisting of two Electromyography (EMG) sensors, two IMU's and a Galvanic Skin Response (GSR) sensor was connected to Arduino 1. IMUs located under the chair seat and on the chair back were managed by Arduino 2. The particle sensor was operated by Arduino 3. Environmental and CO$_2$ sensors were connected to Arduino 4.
Gaming PCs logged data from the eye tracker, keyboard, and mouse. Smartphones collected data obtained by thermal cameras and heart rate monitors.

We synchronized all logging devices using an NTP server. Considering the network delay and time desynchronization during the experiment, we estimate the maximum time divergence as 500 ms.

\begin{figure}[!t]
    \centering
	\centerline{\includegraphics[width=0.8\linewidth]{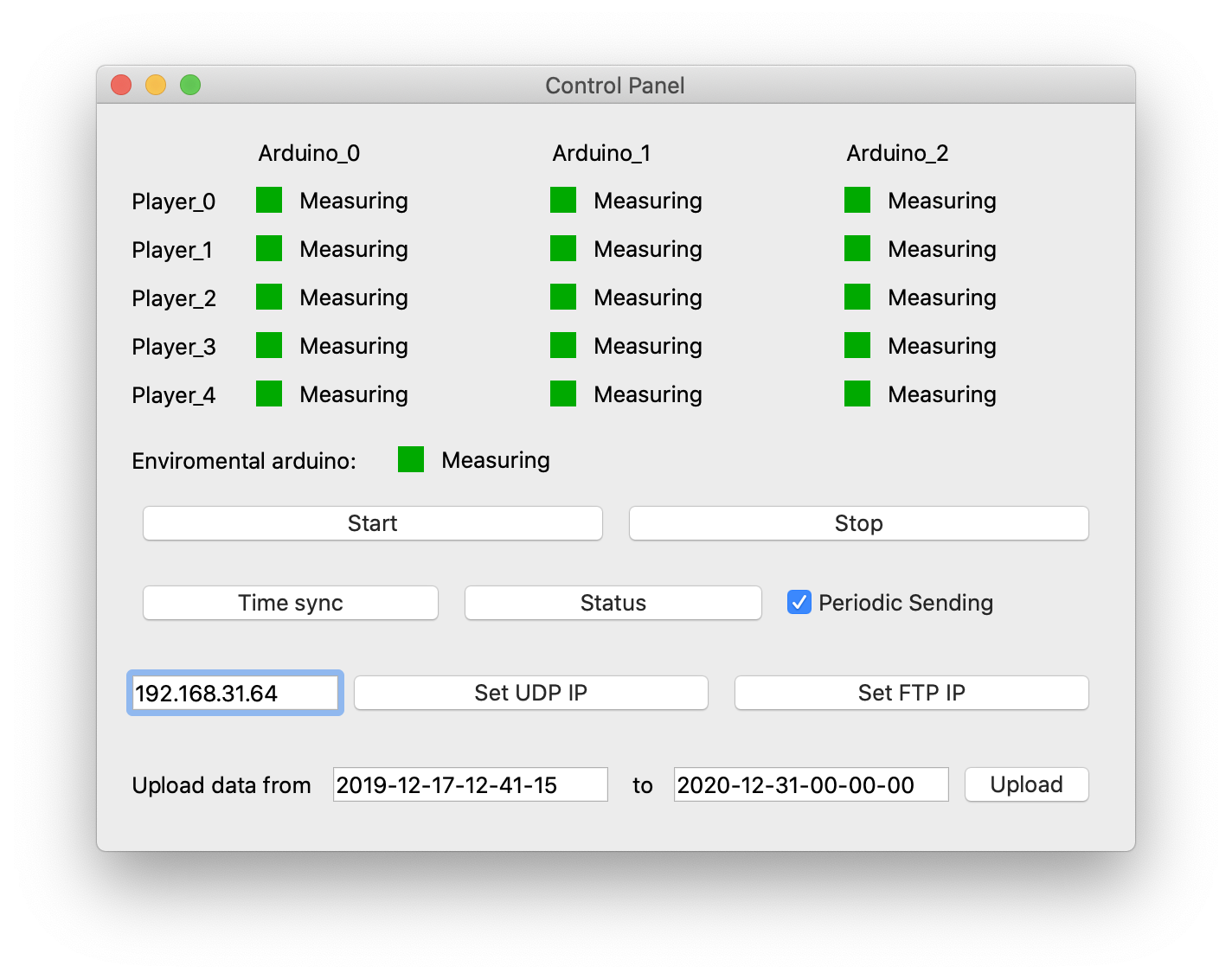}}
	\caption{A control panel used for sensors' time synchronization.}
	\label{control_panel}
\end{figure}

We constructed five copies of a sensing system for simultaneous data collection for five players.
To ensure the synchronization of sensor data, we developed a control panel, which allowed us to start and end measurements simultaneously, synchronize time using NTP server, and monitor the status of measurements. The interface of the control panel is shown in Fig.~\ref{control_panel}. Devices were communicated via UDP protocol over the local Wi-Fi network.

\begin{figure}[!t]
    \centering
	\centerline{\includegraphics[width=1.0\linewidth]{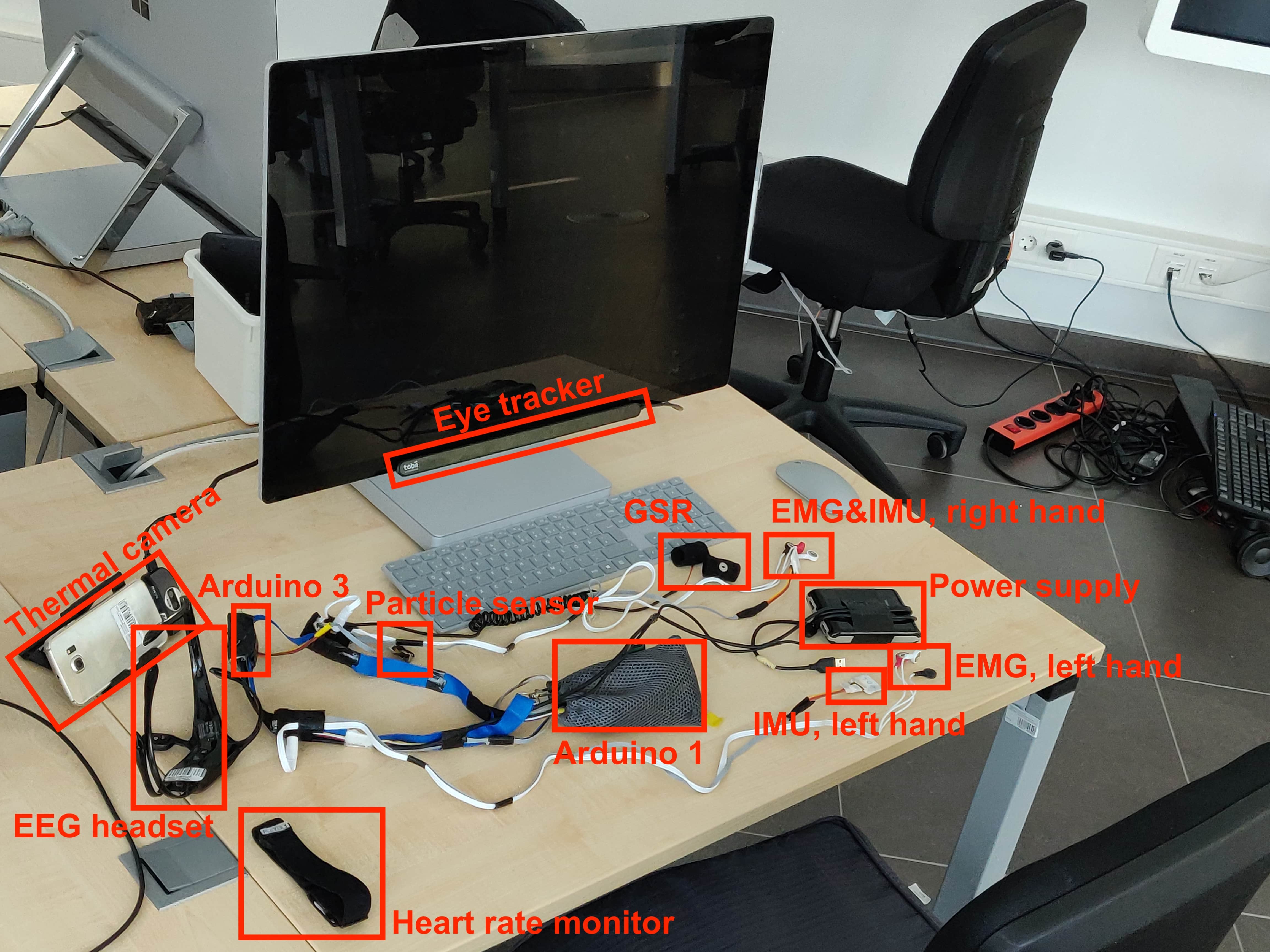}}
	\caption{Experimental testbed.}
	\label{exp_testbed_0}
\end{figure}

\begin{figure}[!t]
    \centering
	\centerline{\includegraphics[width=\linewidth]{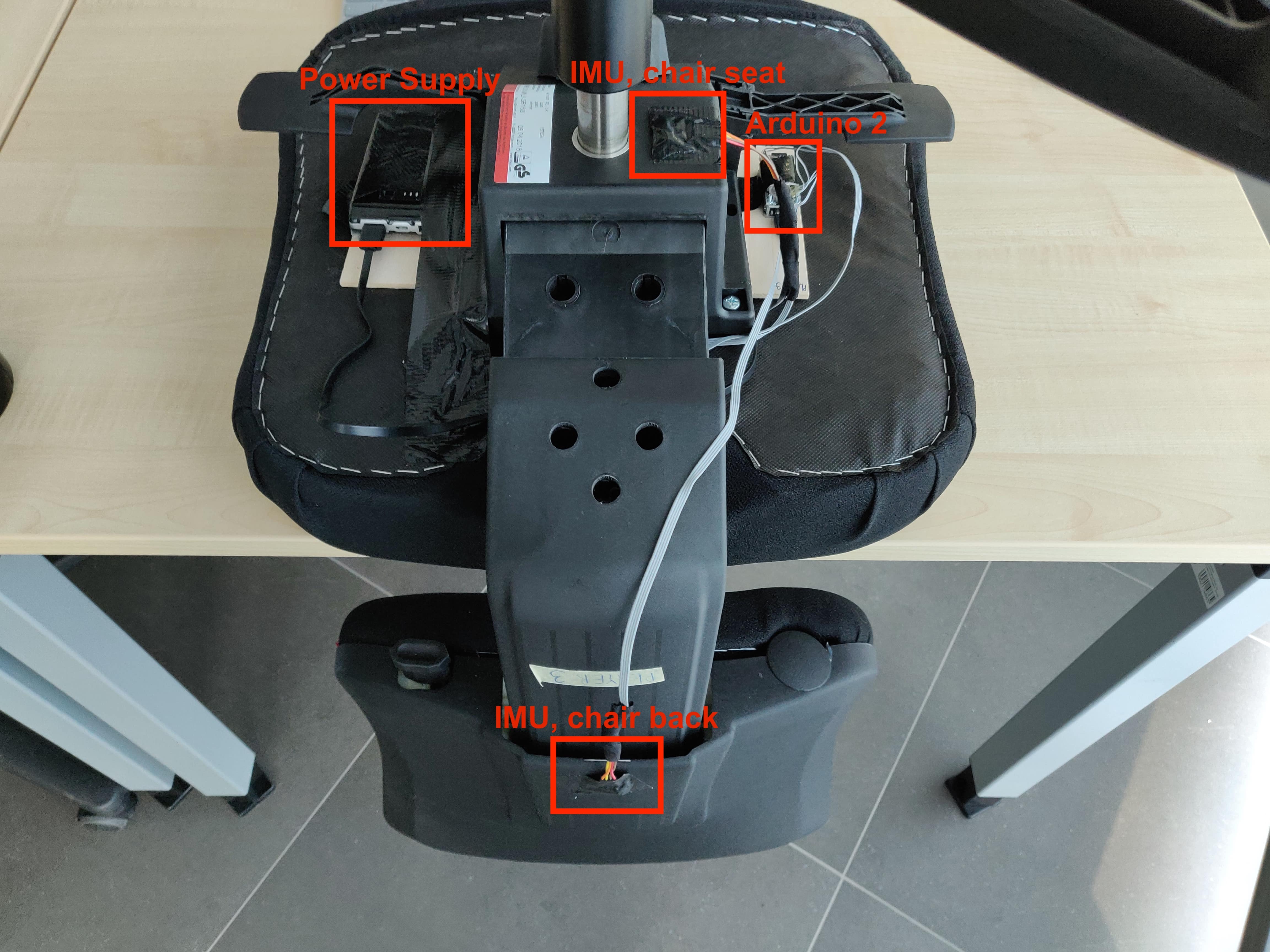}}
	\caption{Sensors integrated into a chair.}
	\label{chair_sensors}
\end{figure}


A wide range of sensors allowed us to capture various indicators of players physiology, including:

\begin{itemize}

\item Lower arm muscle activity as a signal from EMG sensors. It corresponds to intensities of mouse clicks, keystrokes, and general hand movements. EMG data are also correspond to physical tension of players \cite{emg_tension}.

\item Additional data about hand movements obtained from IMUs located on both hands.

\item Person arousal as a skin resistance captured by the GSR sensor \cite{gsr_arousal}.

\item Pulse data collected by heart rate monitor on the upper arm. High pulse implies mental stress and arousal \cite{heart_rate_stress}.

\item Saccades and eye movement activity recorded by the eye tracker. That provides evidence on how players process visual information in the game, such as a minimap, players positioning, etc.

\item Players' activity on a chair as data from IMU sensors located under the chair seat and back. Too intense or frequent movement might be related to high levels of stress \cite{chair_stress}.

\item Facial skin temperature captured by the infrared camera as an indicator of mental load and concentration \cite{face_temperature_mental_load}.

\item EEG signal as a descriptor of brain activity. EEG headsets also include an IMU providing information about players' head movements.

\item Environmental conditions as relative humidity, CO$_2$ level, and temperature. Uncomfortable environmental conditions were shown to affect human cognitive abilities \cite{co2_performance, temperature_performance, humidity_performance}.

\end{itemize}

The implementation of the sensing system is shown in Fig.~\ref{exp_testbed_0} and Fig.~\ref{chair_sensors}.

\subsection{Data Collection Methodology} \label{methodology}

\begin{figure}[!b]
    \centering
	\centerline{\includegraphics[width=0.9\linewidth]{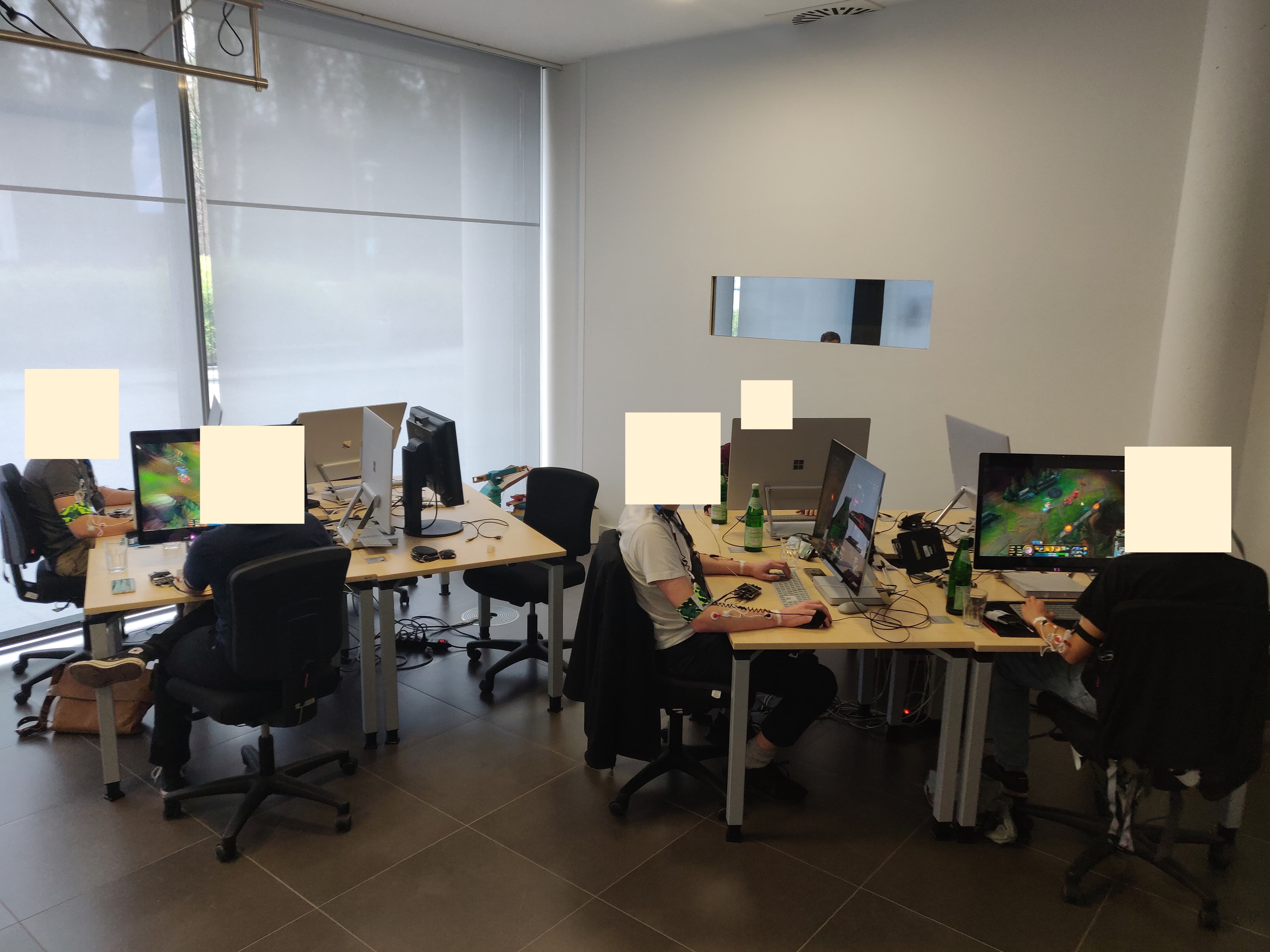}}
	\caption{Data collection process.}
	\label{exp_process}
\end{figure}

We invited a professional and an amateur team to participate in the experiments. Each team played 11 matches in League of Legends over 3 experimental days. 
A process of data collection is shown in Fig.~\ref{exp_process}. Participants were located in the same room and played up to 4 matches on each experimental day with short breaks.

\begin{figure}[!t]
    \centering
	\centerline{\includegraphics[width=1.0\linewidth]{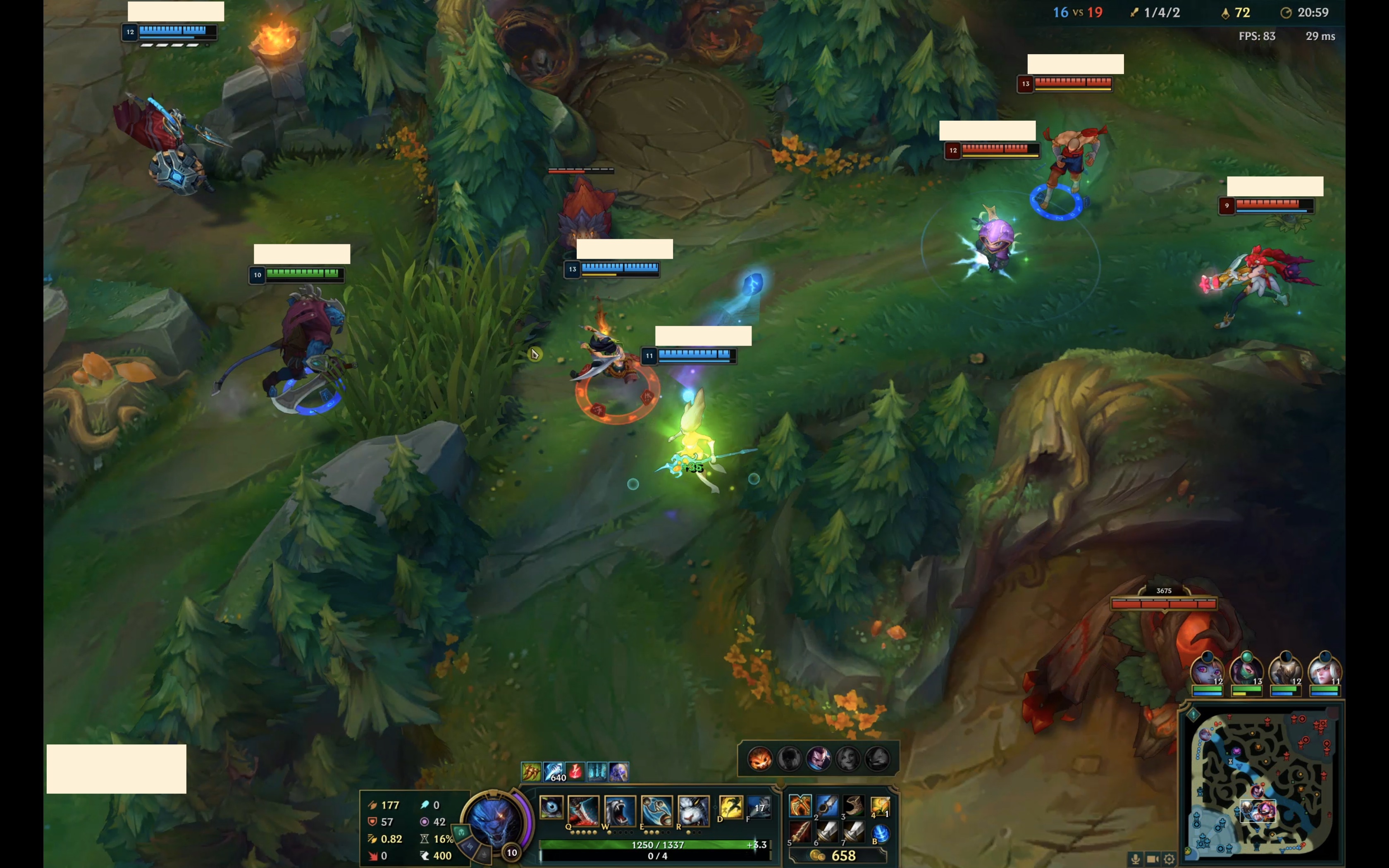}}
	\caption{Game interface.}
	\label{game_interface}
\end{figure}

Game client versions varied from 9.22.296.5720  to  9.24.300.6382 since matches were played on several experimental days.
A typical game interface is presented in Fig.~\ref{game_interface}.

After each match, participants reported if sensing system disturbed them during the game.
In 17.5\% of responses, players weren't concerned about the sensing system; 69.7\% reports indicate a slight discomfort,
12.8\% of records indicate a considerable disturbance caused by the sensing system.

\subsection{In-game Data Collection}

To obtain information about game events, we used API provided by Riot\footnote{\url{https://developer.riotgames.com/}} (League of Legends developer). For each match, we obtained a sequence of in-game events. Typical events parsed are CHAMPION\_KILL, BUILDING\_KILL, SKILL\_LEVEL\_UP, WARD\_PLACED, ITEM\_PURCHASED, etc. The main event we used is CHAMPION\_KILL, which indicates a character being killed, killing and assisting characters, a timestamp, and a location.

\section{Data Analysis} \label{data_analysis}

To extract a meaningful signal from the data, we preprocessed both sensor data and in-game data. The data utilized is available at \url{https://github.com/smerdov/eSports_Sensors_Dataset}.

\subsection{Sensor Data Preprocessing} \label{sensor_data_preprocessing}

Some sensors don't provide a signal as a time series (e.g. mouse, keyboard) or provide uninterpretable raw data (e.g. EMG, GSR).
To convert these data into a meaningful signal, we used signal preprocessing described in Table~\ref{feature_preprocessing}.
Raw signals from some sensors (e.g. heart rate, CO$_2$ level, EEG band power) are directly interpretable, so we didn't modify it at this point.

\begin{table*}[!b]
\centering
\caption{Sensor data preprocessing.}
\begin{tabular}{|p{2.9cm}|p{3.8cm}|p{9.5cm}|}
\hline
Sensor                                        & Raw Data                                    & Processed Data \\ \hline
EMG                                           & Voltage measured by electrodes.             & We calculated muscle activities for both hands as absolute deviations from the reference levels. References levels were calculated as median values. \\ \hline
IMU (right\&left hand, chair seat\&back, head) & Linear acceleration, angular velocity, etc. & The sensors are not calibrated perfectly, so we subtracted the reference level calculated as a median value for linear acceleration and angular velocity and used absolute values of the signal as an indicator of movement activity. \\ \hline
Mouse                                         & Click events, mouse position events.         & The number of mouse clicks in the last 5 seconds, the total distance traveled in the last 5 seconds. \\ \hline
Keyboard                                      & Key pressed/released events.                 & The number of buttons pressed in the last 5 seconds. \\ \hline
Eye tracker                                   & Gaze position in $(x, y)$ coordinates.       & The total distance passed by gaze in the last 5 seconds and pupil diameter. \\ \hline
Facial skin temperature                              & 48x64 heatmap.                             & 95-th percentile of values in the heatmap as an estimation of facial skin temperature. \\ \hline
Particle sensor & The intensity of infrared and red lights passed through the earlobe. & Oxygen saturation (SpO2). \\ \hline
\end{tabular}
\label{feature_preprocessing}
\end{table*}

After all sensor data were converted to time series, 
we resampled the data to a common \textit{time step} $\dt$. All values inside $\dt$ intervals were averaged; if there aren't any values in the interval we used the last available value. $\dt$ is an important hyperparameter determining the fidelity of data representation and indicating how often new data are available to predictive models.
In our work we used $\dt=1$ second.


Sensor data include a lot of noise, which can destabilize machine learning algorithms. To reduce the impact of noise in our models, we got rid of outliers by clipping a signal between 0.05 and 0.995 quantiles.
We also smoothed data from all sensors with an exponential moving average \cite{ema_halflife} using 5 seconds as a half-life interval.

\begin{figure*}[!t]
    \centering
	\centerline{\includegraphics[width=\linewidth]{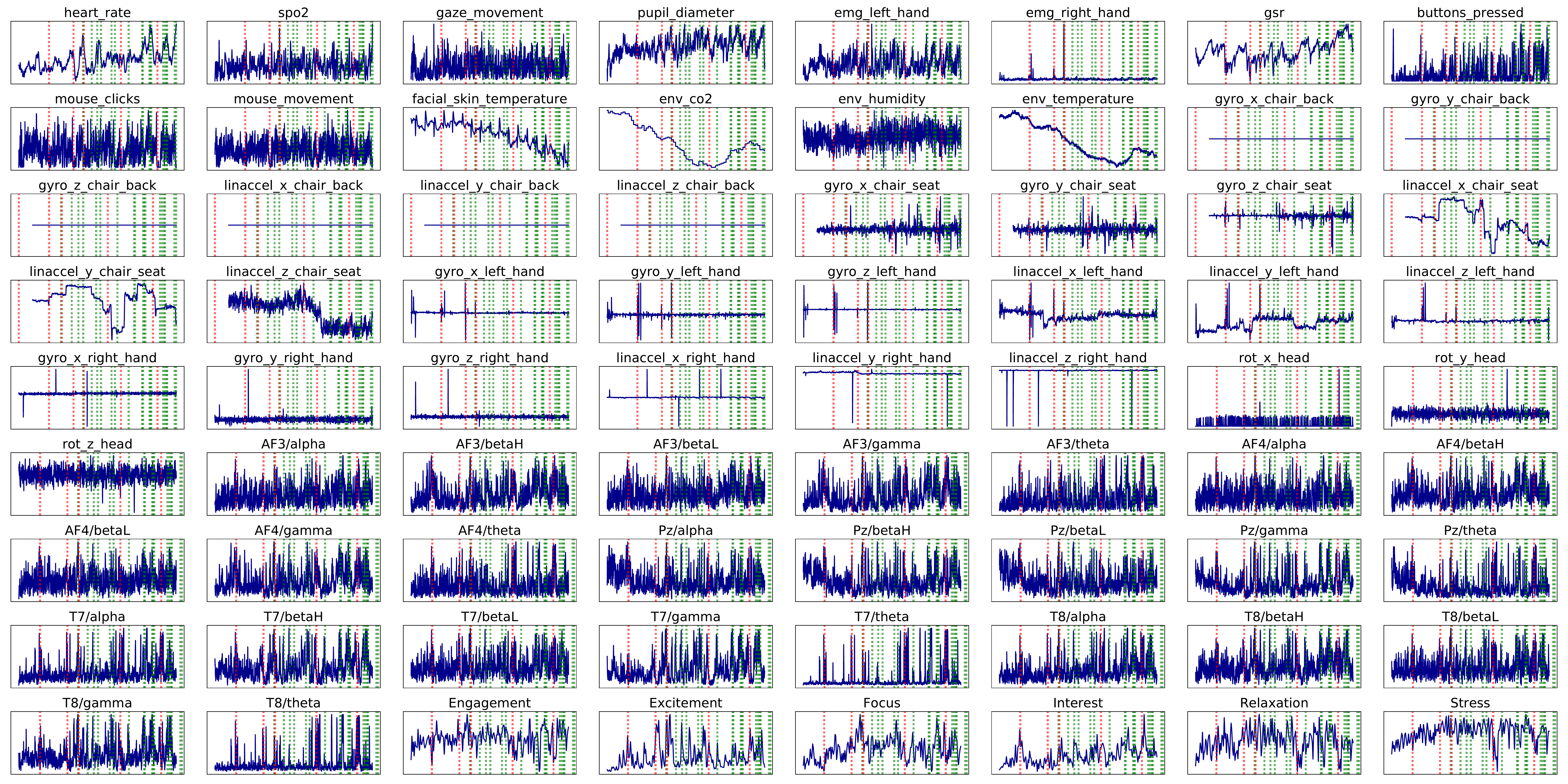}}
	\caption{Sensor data collected for one player in one match. Green and red vertical lines correspond to encounters with positive and negative outcomes, respectively.}
	\label{sensor_data}
\end{figure*}

The resulting signal obtained is shown in Fig.~\ref{sensor_data}. Vertical lines correspond to positive and negative player's impact in encounters, which are described in Subsection~\ref{encounters}.

After preprocessing, we get a sequence of 72-dimensional feature vectors $\xx(t) \in \RRR^{72}$.

\subsection{In-game Data Preprocessing} \label{encounters}

An encounter is an event when two or more heroes from opposite teams are fighting with each other.
The key idea is that players may perform well in the encounter even if their characters die and don't make any direct kills.
Instead, players may successfully assist allied heroes in the fight by contributing healing or damage and don't kill any enemy.
To deal with this problem, 
we propose to calculate a player's impact on the encounter as a cumulative sum of positive and negative in-game events normalized by the number of players participating in the encounter. That metric should properly score players with supportive roles since they aren't making many kills in the game.


From parsed in-game logs, we have a sorted sequence of game event $\big\{e(t_j)\big\}$ 
 for each player, where $t\in\RRR_+$ is a timestamp. An event $e(t)$ corresponds to either kill, death or assist.
For an event occurring at moment $t$, we denote $N(t) \in \NNN$ the number of players participating in the event as killers or assistants, and $i(t)\in \RRR$ player's impact in the event.
It is reasonable to assume that player's impact is positive in kill/assist events and negative in death events. Also, it's rational to 
normalize player's impact in the event by the number of participating players, so we used the following formula to calculate the player's impact:

\begin{align}
i(t) =
  \begin{cases}
    \frac{+1}{N(t)}&, \text{~~if event $e(t)$ is kill or assist}\\
    -1&, \text{~~if event $e(t)$ is death}
  \end{cases}
  \label{event_impact}
\end{align}

Equation \ref{event_impact} ensures that the total impact of all players sums up to zero.

Normalization by a number of participating heroes ensures that a player's impact is not overestimated when she makes a kill (or 'frag' in the players slang) as a part of a team. 
However, it's incorrect to estimate player performance by a single event only. We propose to aggregate player impact over a
sequence of adjacent events, or encounter, to estimate her performance more precisely.

In League of Legends, an encounter typically consists of several events (kill, death, assist) occurring in a short period of time, while the interval between encounters is much longer.
We propose to group events into encounters
such that the interval between any events from different encounters exceeds some minimum interval $T$.

That defines a sequence of encounters $\big\{E(t_j)\big\}$, 
where each encounter 
$E(t_j) = \big\{e(t_{j}), e(t_{j+1}), ..., e(t_{{j+k_j}})\big\}$ consists of $k_j$ events, such that $0 \leq t_{i+1} - t_i \leq T \ \forall i \in \{j+1,\ldots,j+k_j-1\}$, and also the interval between events from different encounters are more than $T$.
We denote the timestamp of the encounter as a timestamp of the first event in it. We used $T=10$ seconds as a reasonable minimum interval between two encounters.



To calculate a player's impact in each encounter, we summarized all impacts for the events in the encounter:

\begin{equation}
    I(t) = \sum_{i(t') \in E(t)} i(t'),
\end{equation}
where $I(t)$ is the player's impact for the encounter started at moment $t$.
The impact $I(t)$ is positive when the player performs well and is negative when the player plays poorly.


$I(t)$ is a proper metric for performance evaluation: positive values mean that player contributed to kills or assists events and probably didn't die.
In opposite, negative values mean that played died in the encounter and didn't contribute much to the fight.
If $I(t)$ equals 0, we reassign it to the sign of the first event in the encounter.

However, it's more convenient to work with it as a binary value for better interpretability. We binarized the target value:

\begin{equation}
    \tilde{y}(t) = \text{sign}(I(t)),
\end{equation}
where $\tilde{y}(t)$ equals 1 if player performed well in the encounter at timestep $t$, and equals $-1$ otherwise.

Encounter outcome suppose to estimate players performance better than classical metrics like Kill Death Ratio \cite{kdr}, since these metrics underestimate players in supportive roles and don't utilize the encounter-like nature of game events.

A natural idea is to predict whether player will perform good in the next encounter by sensor data. We use \textit{forecasting horizon} hyperparameter $\tau$ to define how much time before the encounter we are trying to predict how player will perform.
That leads us to the final target value:

\begin{equation}
\yy(t) = \tilde{y}(t + \tau),
\end{equation}
where $\yy(t)$ is a target indicating if the player will have a good impact in the fight occurring $\tau$ seconds later.

\subsection{Dataset and Evaluation} \label{dataset_and_evaluation}


Since both teams participated in 3 experimental days, we used data from the first two experimental days for training and data from the third experimental day for testing.

As input data for machine learning algorithms, we used sequences of 72-dimensional feature vectors
(see Subsection~\ref{sensor_data_preprocessing})
for each match and player. As a target, we utilized $\yy(t)$, indicating whether a player will have a positive or negative impact in the encounter $\tau$ seconds later
(see Subsection~\ref{encounters}).

To ensure proper scaling of data for machine learning methods, we standardized data for each player, match, and feature to zero mean and unit variance. Missed values were filled using linear interpolation or median values.

The total number of encounters in the dataset is 888. 56.1\% of them belong to the positive class; 43.9\% belong to the negative class.
We used ROC AUC \cite{roc_auc} as an evaluation metric as it properly suits a balanced binary classification task. It ranges from 0 to 1 and achieves a 0.5 score for random guess (even for unbalanced data). Higher values are better.
ROC AUC demonstrates the diagnostic ability of the classifier with varying thresholds.

At first, we calculated scores for each player and then averaged them for all participants. That ensures that models are evaluated on how they estimate players' performances instead of separating newbies and experienced players.

\section{Machine Learning} \label{machine_learning}

\subsection{Methods Overview}

We utilized several models for prediction of encounter outcomes including reasonably simple and well-known methods, e.g. logistic regression, k-nearest neighbors (KNN), as well as more complex and novel models, e.g. gated recurrent unit (GRU), transformer.
Some of the methods (GRU, transformer) are designed for Seq2Seq problems like ours, while others (Logistic regression, KNN) do not take the sequential nature of data into account. All these models use sensor data as input.
To check if sensor data is helpful for prediction, we compared above mentioned methods with the baseline model, which did not utilize sensor data.

\begin{figure*}[!b]
    \centering
	\centerline{\includegraphics[width=0.85\linewidth]{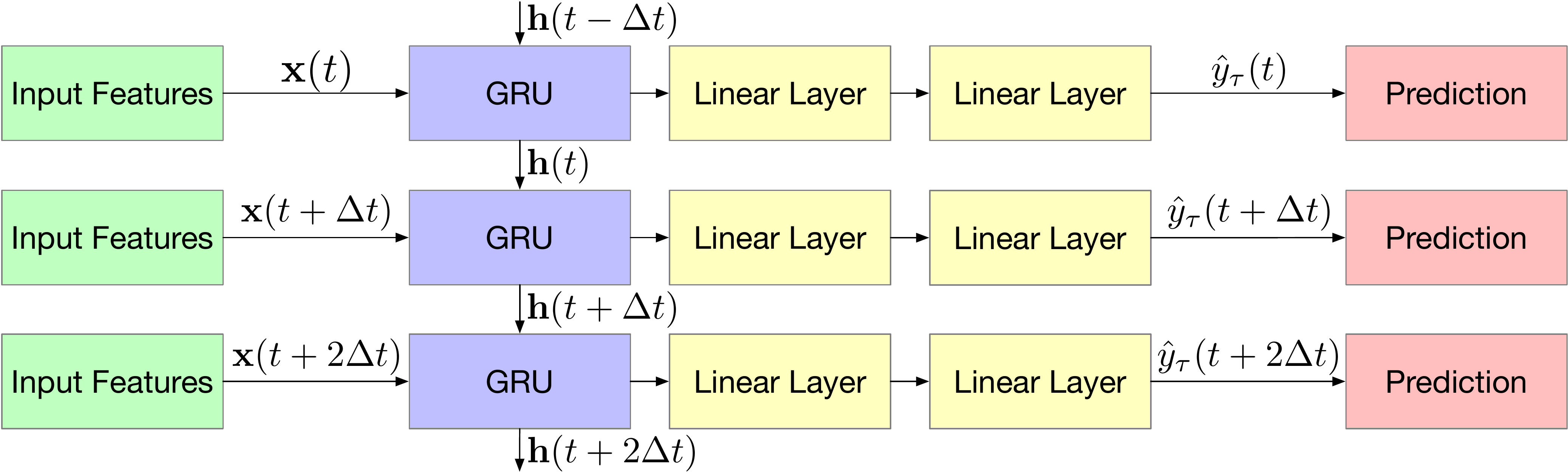}}
	\caption{Gated recurrent unit network architecture.}  
	\label{gru_arch}
\end{figure*}

\begin{figure*}[!t]
    \centering
	\centerline{\includegraphics[width=0.85\linewidth]{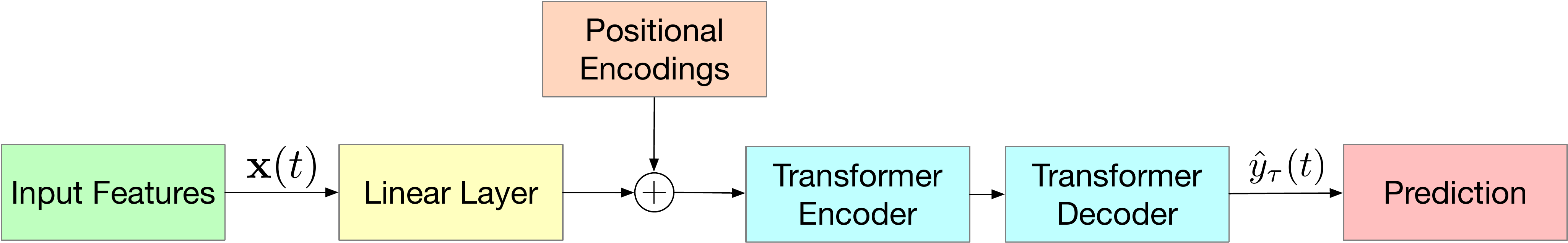}}
	\caption{Transformer network architecture.}  
	\label{transformer_arch}
\end{figure*}

\begin{enumerate}
\item \textbf{Baseline model.} This model uses the last observed target value as a prediction.
For example, it predicts $\hat{y}_\tau(t)=1$ if at the moment $t$ the latest observed target was positive, and $\hat{y}_\tau(t)=0$ if at the moment $t$ the latest observed target was negative.
In other words, the model predicts the last seen target value for the encounter occurring $\tau$ seconds later.

The baseline model doesn't utilize sensor data, and this allows us to check how access to sensor data could improve the quality of predictions.
Also, this model is the only model utilizing target values during the inference.

\item \textbf{Logistic Regression.} A simple linear method for data classification \cite{logistic_regression}. It outputs a probability that a feature vector $\xx(t) \in \RRR^n$ belongs to the positive class:

\begin{equation}
P\Big(\yy(t) = 1\Big|\xx(t)\Big) = \frac1{1 + \exp\Big(-\big\langle \ww, \xx(t) \big\rangle + b\Big)},
\end{equation}

\item \textbf{KNN}, or k-nearest neighbors algorithm for data classification \cite{knn}. It makes a prediction based on a similarity in a feature space. We found $k=96$ to be the optimal hyperparameter in our experiments.

\item \textbf{Gated recurrent unit (GRU) neural network} \cite{gru}. This is a type of recurrent neural network utilizing gating mechanism for better information flow. More specifically, GRU incorporates update gate $\zz(t)$ and reset gate $\rr(t)$. The update gate controls how new input changes the network hidden state $\hh(t)$, while the reset gate is responsible for keeping or resetting the previous value of the hidden state. Formally:

\begin{gather}
\hh(t) = \big(1 - \zz(t)\big) \odot \hh(t-\dt) + \zz(t) \odot \tilde{\hh}(t),\\
\tilde{\hh}(t) = \tanh\Big(\WW_h \xx(t) + \UU_h \big(\rr(t) \odot \hh(t-\dt)\big) + \bb_h\Big),\\
\zz(t) = \sigma\big(\WW_z \xx(t) + \UU_z \hh(t-\dt) + \bb_z\big),\\
\rr(t) = \sigma\big(\WW_r \xx(t) + \UU_r \hh(t-\dt) + \bb_r\big),
\end{gather}
where $\WW_h, \WW_z, \WW_r, \UU_h, \UU_z, \UU_r, \bb_h, \bb_z, \bb_r$ are the learnable parameters, $\sigma$ is the sigmoid function, $\odot$ is Hadamard product.

GRU is similar to LSTM, but it benefits from a fewer number of parameters. In our experiments, GRU performed better than LSTM. In our experiments the best GRU network had one recurrent layer with hidden size 64 and two subsequent linear layers. The GRU architecture is shown in Fig.~\ref{gru_arch}. We used ReLU activation for linear layers \cite{relu} and sigmoid activation for the network output.

\item \textbf{Transformer neural network.} It was proposed in \cite{att_all_you_need} to better deal with problems in natural language processing, but it was found to work well in time series forecasting \cite{transformer_timeseries}.
The transformer architecture eschews recurrence and instead
relies entirely on an attention mechanism to draw global dependencies between input and output.
To utilize the sequential nature of data, transformer architecture utilized positional encoding:

\begin{align}
    PE_{(pos, 2i)} &= \sin \big(pos/10000^{2i / d}\big),\\
    PE_{(pos, 2i+1)} &= \cos \big(pos/10000^{2i / d}\big),
\end{align}
where $pos$ is the position, $i$ is the dimension, $d$ is the number of embedding dimensions (we used $d=64$). Positional encodings are added to embeddings of sensor data to represent the current timestamp.

The transformer also incorporates stacked encoder-decoder architecture for high-level feature representation, 
position encoding for capturing the spatial location of inputs,
and multi-head attention blocks for focusing on the most important features.
In our experiments, we used a transformer model with 1 layer for encoder and decoder,
8 heads for multi-head attention, size of fully-connected layer equals 512, and dropout equals 0.5.
To convert sensor data into proper embeddings, we added a 64-dimensional linear layer before the transformer encoder.
Outputs of the transformer decoder were normalized to [0,1] interval with sigmoid function to represent probability.

The architecture of the transformer network is shown in Fig.~\ref{transformer_arch}.

\end{enumerate}

\begin{figure*}[!b]
    \centering
	\centerline{\includegraphics[width=0.8\linewidth]{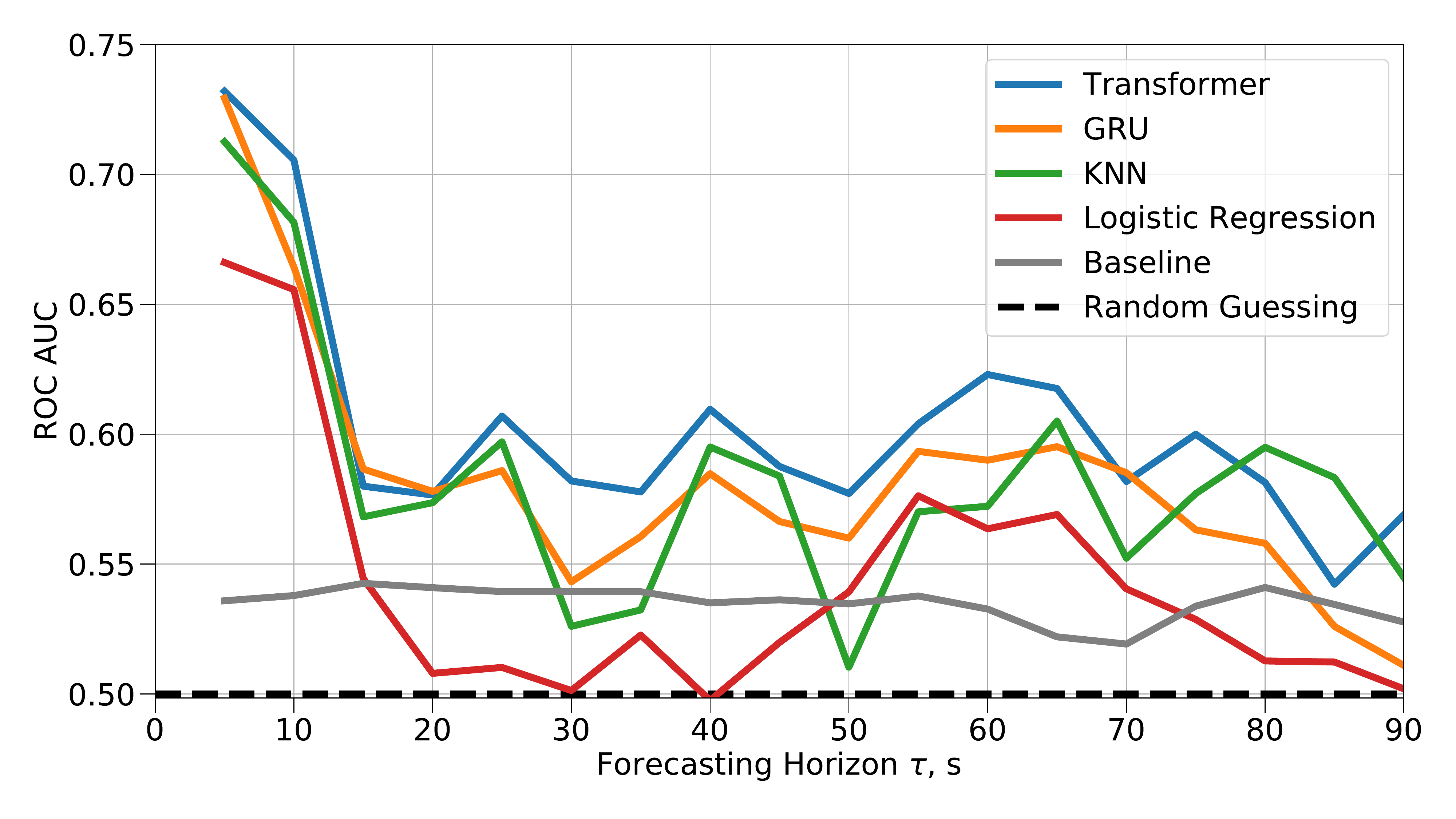}}
	\caption{Comparison of different algorithms w.r.t. forecasting horizon.}
	\label{scores_forecasting_horizon}
\end{figure*}

We used \texttt{scikit-learn}\footnote{\url{https://scikit-learn.org/}} python library for logistic regression and KNN implementations. For GRU and transformer networks, we utilized \texttt{pytorch}\footnote{\url{https://pytorch.org/}}. Networks were optimized by Adam optimizer \cite{adam} with standard parameters ($\alpha=0.001, \beta_1 = 0.9, \beta_2=0.999, \epsilon=10^{-8}$).

\begin{figure}[!t]
    \centering
	\centerline{\includegraphics[width=\linewidth]{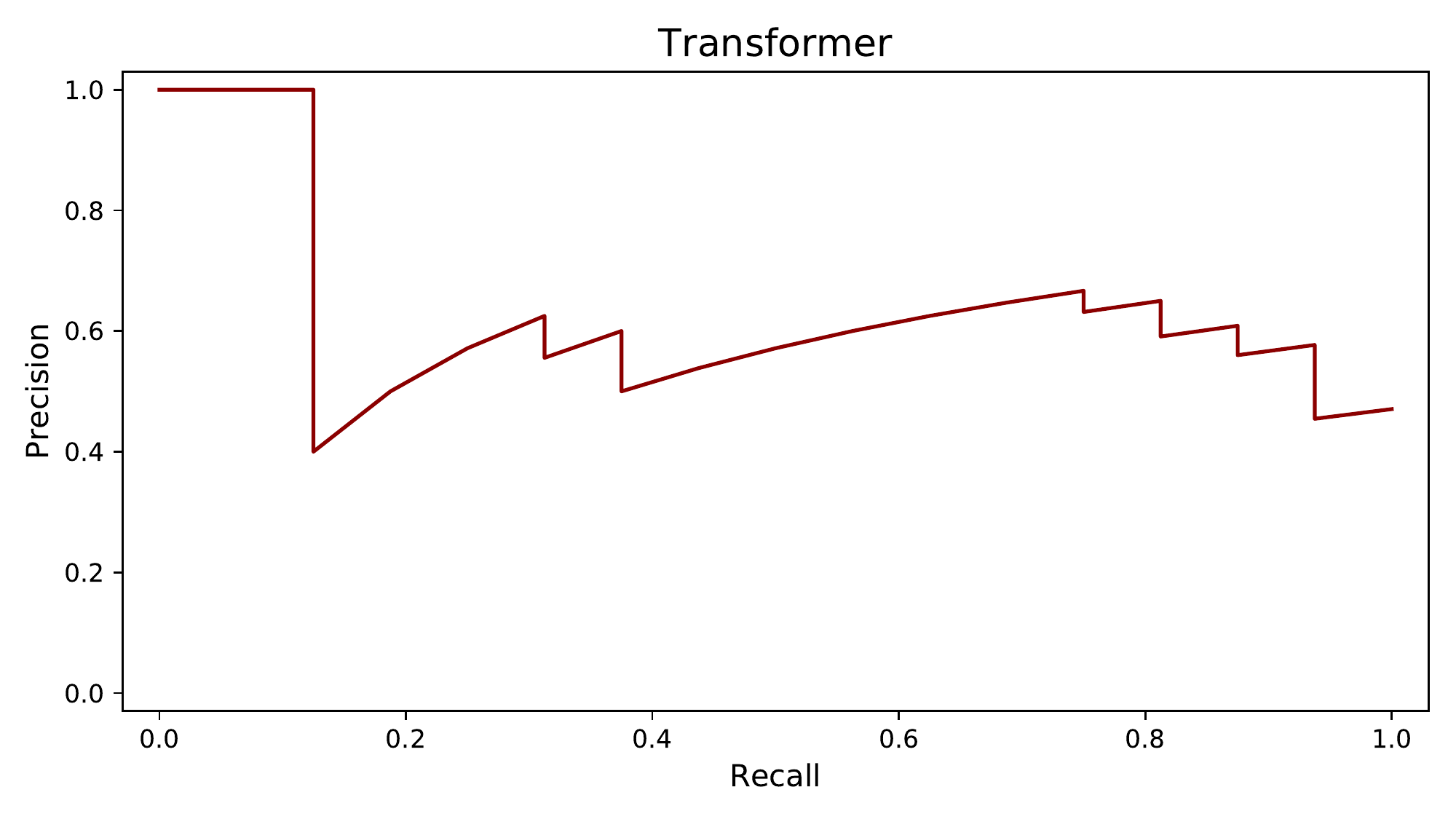}}
	\caption{A sample of precision-recall curve for a transformer model for one player and forecasting horizon $\tau=10$ seconds. Here we used encounter loss as a positive class.}
	\label{precision_recall}
\end{figure}

\begin{figure*}[!t]
    \centering
	\centerline{\includegraphics[width=0.9\linewidth]{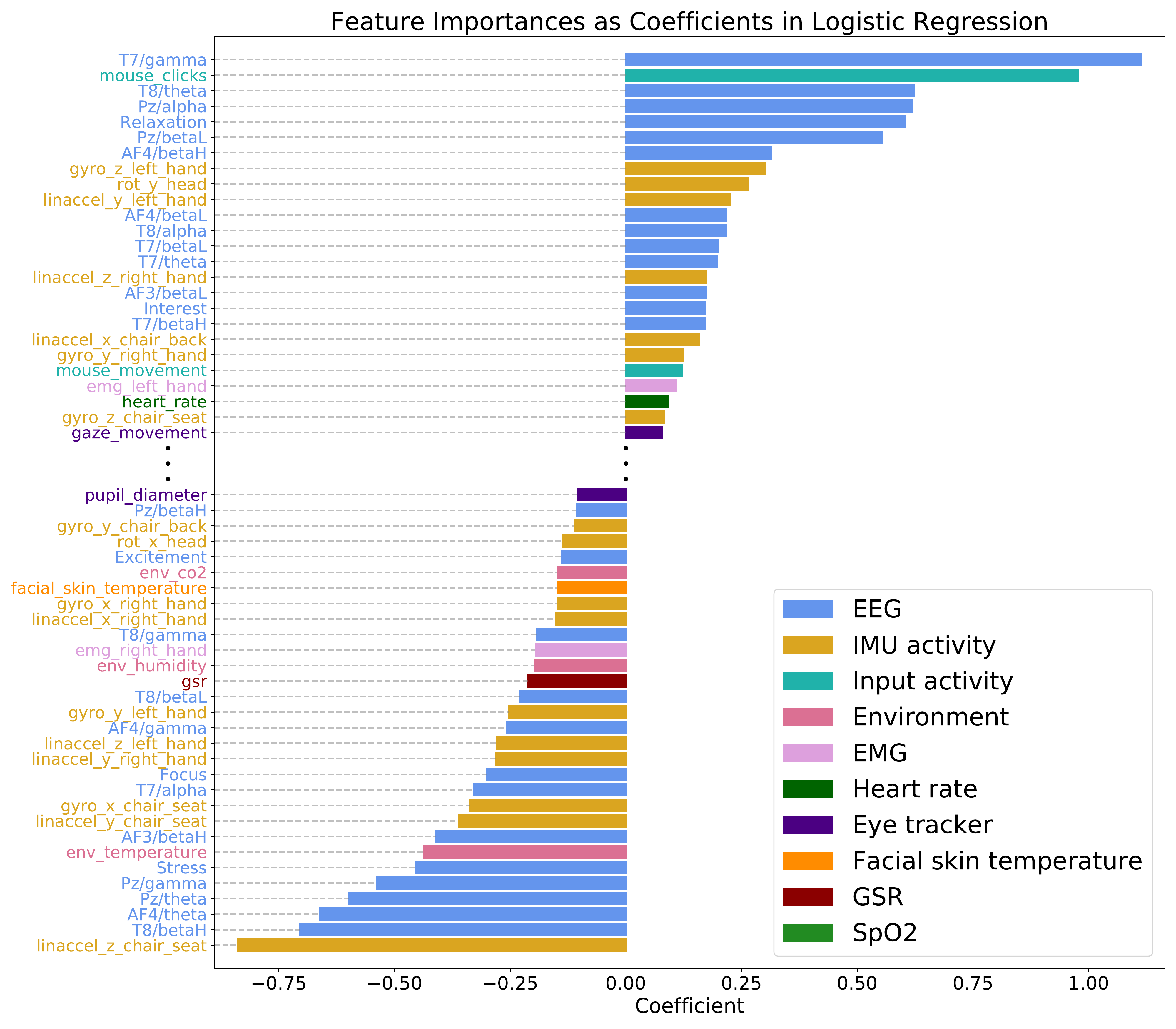}}
	\caption{
	Feature importances as coefficients in logistic regression model. Positive coefficients correspond to features having higher values 10 seconds before the successful fight; the opposite holds for negative coefficients.
	Features are colored with respect to sensors.
	The figure demonstrates the top-25 most positive and the top-30 most negative coefficients out of 72 features.}
	\label{feature_importance}
\end{figure*}

\subsection{Results}

We have trained the above-mentioned machine learning models on the dataset described in Subsection~\ref{dataset_and_evaluation} for varying forecasting horizons $\tau$. The results are illustrated in Fig.~\ref{scores_forecasting_horizon}. We averaged results of 5 reruns for each algorithm.


Short forecasting horizons like 5 or 10 seconds ensure higher scores because models need to predict encounter outcomes in the nearest future. Scores are naturally decreasing when the forecasting horizon increases; however, there is a slight increase in scores near 50-70 seconds. The possible explanation is that the result of the fight occurring after 50-70 seconds might be the same as the fight occurring after 5-10 seconds because a typical champion respawn time in League of Legends is 20-40 seconds, so the next encounter has a high chance to occur in 20-30 seconds after the respawn.

Interestingly, one of the worst-performing models is a baseline model, which doesn't utilize sensor data at all, but uses a history of encounter outcomes instead. 
This might imply that sensor data can provide more insights about the player condition rather than in-game records.

We suppose that $\tau=10$ seconds forecasting horizon is a sweet spot in terms of 
predictive power and potential usage.
We further focus on the analysis of the transformer model, which is the best for this forecasting horizon.

One of the potential applications of trained models is a player's state assessment. If a model predicts that a player is likely to lose the next encounter (i.e., because of burnout, fatigue), it might suggest the player to retreat or to stick to a more defensive strategy. In opposite, if the algorithm predicts that the player is in good form, she might make riskier in-game decisions.

To check the possibility of creating a player's burnout detection system, we checked the precision and recall of the best models. For a transformer and forecasting horizon $\tau = 10$ seconds, a typical precision-recall curve is shown in Fig.~\ref{precision_recall}. We observed similar precision-recall curves for other players.

For each player, we selected a probability threshold value on training data, such that a recall for encounter losses is less or equals 50\%. Then we binarized predictions according to the thresholds estimated for each player. The mean precision of such predictions for all players was 73.5\%, and the mean recall is 88.3\%. In other words, the model managed to detect that a player will lose the next fight 10 seconds in advance in 88.3\% of cases with 73.5\% accuracy. A potential system's user can adjust thresholds to increase recall at the expense of precision, or vice versa.


\subsection{Feature Importance}
Tounderstand attributes influencing player performance
and to
evaluate the contribution of each individual sensor into the resulting model, we calculated feature importances as coefficients in logistic regression. 
High positive coefficients for features imply that these features are generally higher before the encounter win and lower before the encounter loss. In opposite, lesser negative coefficients mean that corresponding features are usually lower before the encounter win and higher before the encounter loss.
More important sensors have more features with high absolute values for coefficients.

We have calculated feature importances for all 72 features and presented the top-25 highest and the top-30 lowest in Fig.~\ref{feature_importance}. The features are colored corresponding to the sensor which obtained this feature.

The feature indicating good in-game activity the most is the intensity of gamma waves for the T7 electrode located near the left temple. The higher intensity of gamma waves is
connected with better cognition performance, selective attention, and conscious perception of visual objects \cite{gamma_eeg}. Prior research in \cite{esports_eeg_0} also showed the increased intensity of theta-waves for positive game events.

Another activity indicating that a player has a higher chance of winning the next encounter is mouse activity: the number of mouse clicks in the last 5 seconds \texttt{mouse\_clicks} and distance traveled \texttt{mouse\_movement} in the last 5 seconds.
Higher heart rate, and eye activity (feature \texttt{gaze\_movement}) also increase the player's performance.
A possible explanation is that
higher heart rate mobilizes players,
and higher eye activity 
demonstrates better awareness of the game screen.

The most important indicator that the player is prone to lose the next fight is a vertical movement of the chair (feature \texttt{linaccel\_z\_chair\_seat}).
This probably corresponds to events when the player changes the posture on the chair or adjusts the chair's height, and then got an unexpected fight. This feature has also been reported by \cite{smart_chair_wf_iot} as the most intrinsic for low-skilled players. Many other activities captured by IMUs are also negatively impacting player performance.

Higher environmental temperature, humidity, and CO$_2$ level decrease the player's performance.
Higher values of facial skin temperature, GSR, and pupil diameter, which are connected with a lower concentration, higher mental load, and stress, lower the probability of winning the future fight.


Interestingly,
higher activity of the left hand (feature \texttt{emg\_left\_hand}) increases the chance to win the encounter, while 
the higher activity of the right hand (feature \texttt{emg\_right\_hand}) decrease this chance.
A possible explanation is that more intensive keystrokes made by the left hand might correspond to shorter reaction time and more abilities used,  while more intensive clicks by the right hand do not change much because one regular click is enough. 





\section{Conclusions and Future Work} \label{conclusion}

In this paper, we have proposed a sensor network architecture for data collection in the eSports domain. We have collected a dataset from professional and amateur players, including sensor data and in-game logs. To extract meaningful data from in-game events, we have proposed a method for automatic extraction and annotation of in-game encounters.
We have trained several machine learning algorithms, including Gated Recurrent Unit and Transformer, to predict encounters' outcomes solely by sensor data. The best model achieves ROC AUC 0.706 in predicting if the player will win the encounter occurring 10 seconds later. We have also converted this model into a players' burnout detection system, which can predict for each player a chance to lose the next encounter with 73.5\% precision and 88.3\% recall. 
The results have been interpreted by feature importance techniques, which allow us to identify physiological features affecting players' performance.

The limitations of this study is a small number of participants involved and a lack of human-labeled annotation for in-game replays. Future work includes more extensive data collection with more subjects recorded, proper data annotation with professional players and coaches, and a diverse set of games analyzed.
These would allow researchers to utilize more complex machine learning models and better understand players' physiological behavior.

\section*{Acknowledgment}
The reported study was partially funded by RFBR (Russian Foundation for Basic Research) according to the research project No. 18-29-22077$\backslash$20.

The research reported in this paper has been partially supported by the BMBF (German Federal Ministry of Education and Research) in the project HeadSense (project number 01IW18001). The support is gratefully acknowledged.

Also, the authors would like to thank the eSports club\footnote{\url{https://www.esports-kl.de/}} at the University of Kaiserslautern for their contribution to methodology development and participation in data collection.




\bibliography{references}{}
\bibliographystyle{IEEEtran}

\vskip -2\baselineskip plus -1fil

\begin{IEEEbiography}[{\includegraphics[width=1.1in,height=1.25in,clip,keepaspectratio]{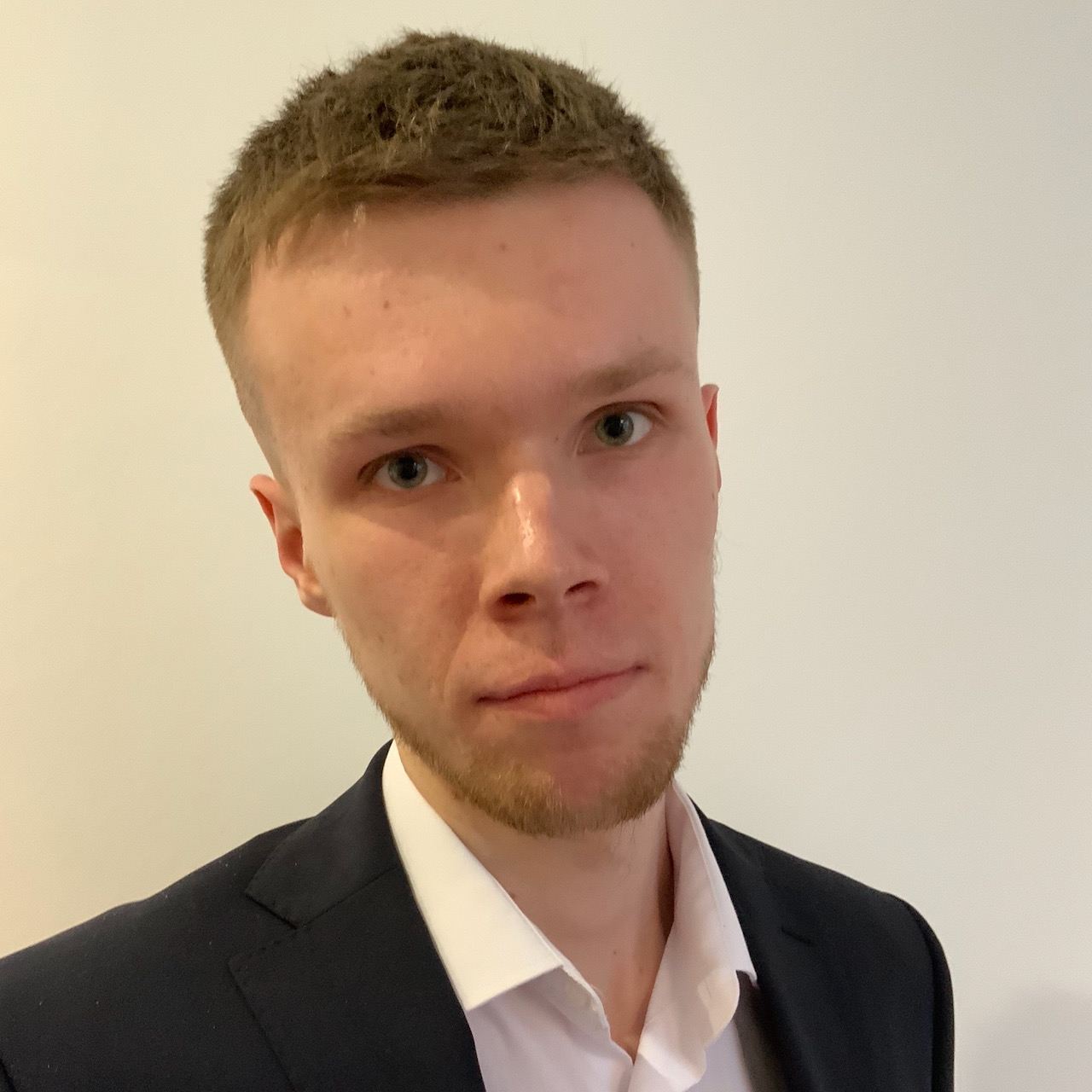}}]{Anton Smerdov} received MS degree in Data Science at Skolkovo Institute of Science and Technology and Moscow Institute of Physics and Technology (MIPT) in 2020, and BS degree in Applied Physics and Mathematics at MIPT in 2018. His research interests include machine learning, artificial intelligence, and computer vision.

\end{IEEEbiography}

\vskip -2\baselineskip plus -1fil

\begin{IEEEbiography}[{\includegraphics[width=1.1in,height=1.25in,clip,keepaspectratio]{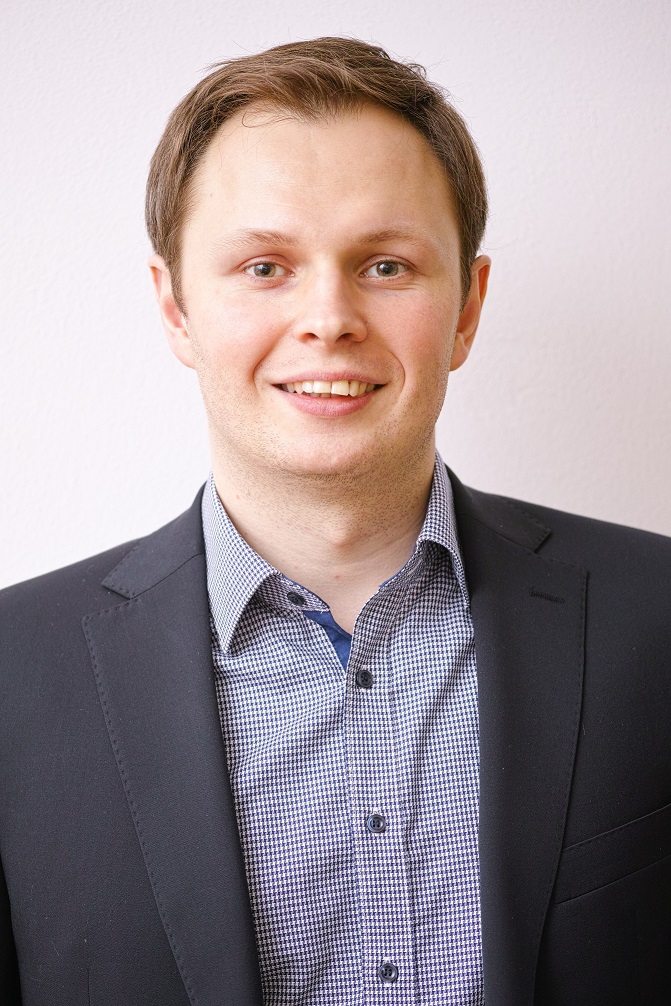}}]{Andrey Somov} is an Assistant Professor at Skolkovo Institute of Science and Technology (Skoltech), Russia. 
Andrey received his PhD (2009) from the University of Trento, Italy, for work in the field of power management in wireless sensor networks (WSN). Before joining Skoltech (2017), he had worked as a Senior Researcher for FBK CREATE-NET Research Center, Italy (2010-2015) and as a Research Fellow for the University of Exeter, UK (2016-2017). Andrey has published more than 80 papers in peer-reviewed international journals and conferences. His current research interests include intelligent sensing, power management for the wireless sensor nodes, cognitive IoT and associated proof-of-concept implementation. Dr. Somov holds some awards in the fields of WSN and IoT including the Google IoT Technology Research Award (2016) and the Best Paper Award at IEEE IoP conference (2019).
\end{IEEEbiography}

\vskip -2\baselineskip plus -1fil

\begin{IEEEbiography}[{\includegraphics[width=1.1in,height=1.25in,clip,keepaspectratio]{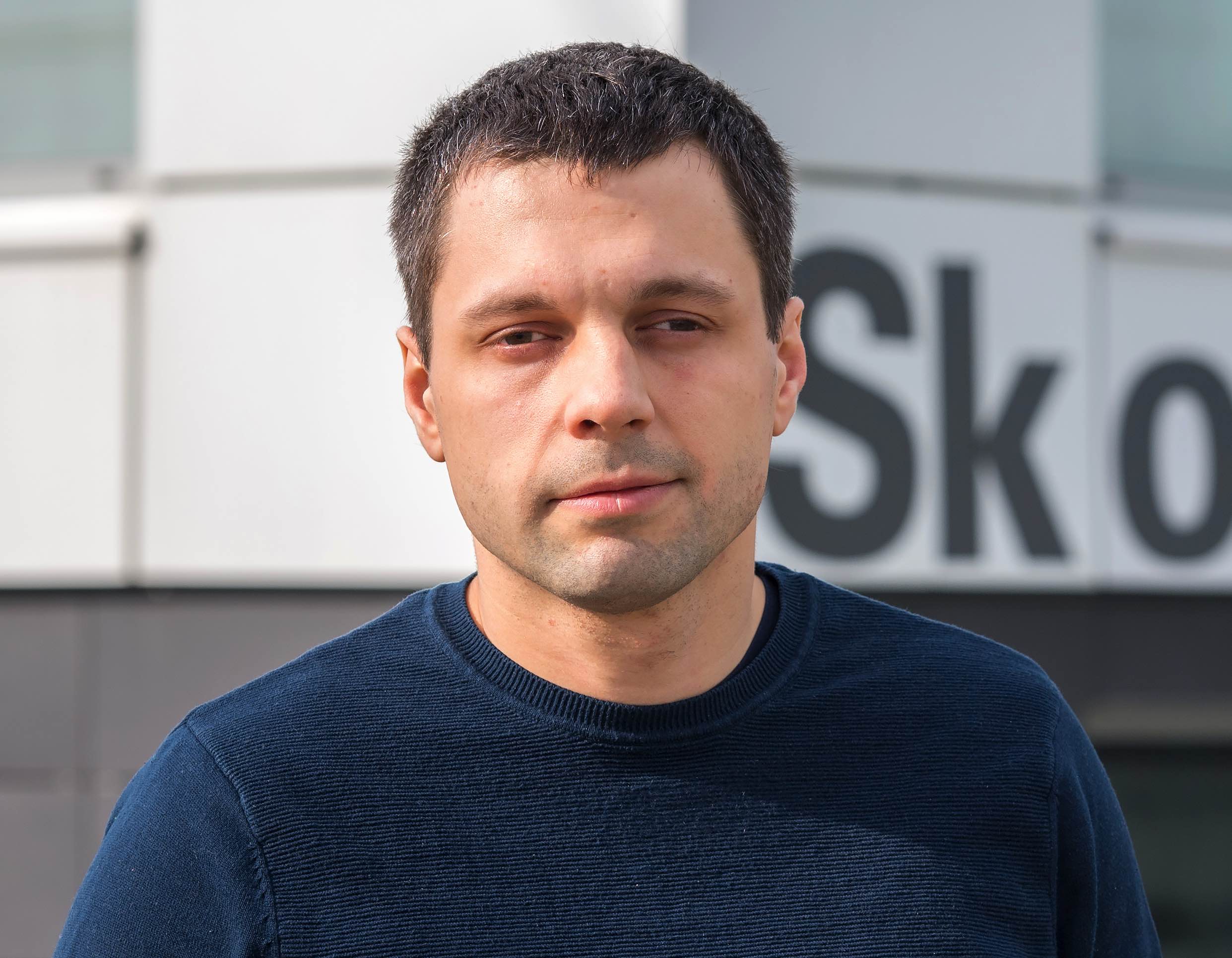}}]{Evgeny Burnaev} is an Associate Professor at Skolkovo Institute of Science and Technology (Skoltech), Russia, and the Head of Advanced Data Analytics in Science and Engineering group at Skoltech. Evgeny’s current research focuses on the development of new algorithms in machine learning and artificial intelligence such as deep networks for an approximation of physical models, generative modeling, and manifold learning, with applications to computer vision and 3D reconstruction, neurovisualization. 

\end{IEEEbiography}

\vskip -2\baselineskip plus -1fil

\begin{IEEEbiography}[{\includegraphics[width=1.1in,height=1.25in,clip,keepaspectratio]{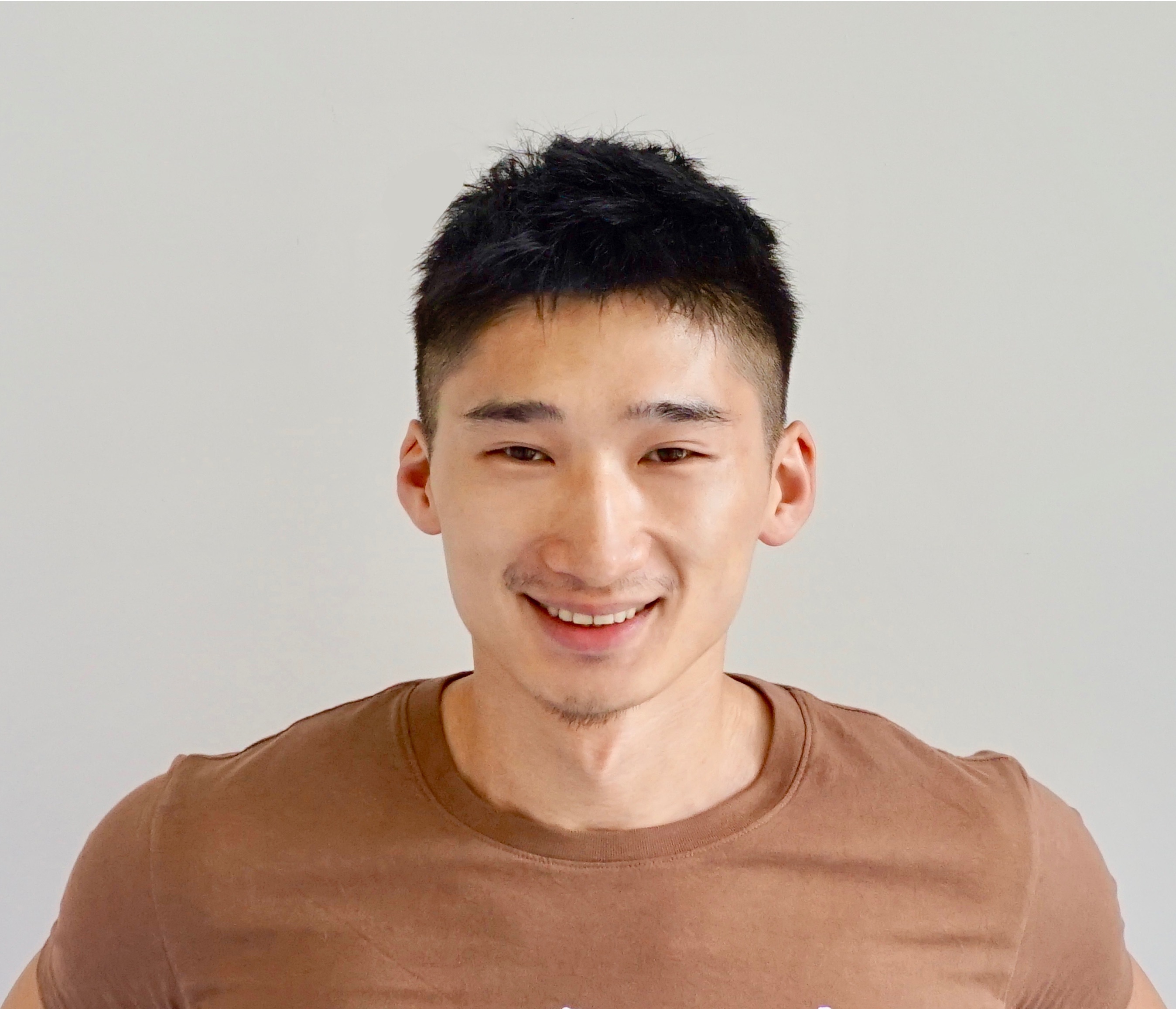}}]{Bo Zhou} is a Senior Researcher at the German Research Center for Artificial Intelligence (DFKI), Germany. He received his PhD (2019) from the University of Kaiserslautern, Germany, for his work in human activity recognition (HAR) with textile pressure mapping (TPM) technologies. Bo has over 30 peer-reviewed academic publications. His current research interests include wearable computing, machine learning with multi-dimensional heterogeneous sensor data, digital sports assistance, biomedical and healthcare engineering, HAR applications of Internet of Things. 

\end{IEEEbiography}

\vskip -2\baselineskip plus -1fil

\begin{IEEEbiography}[{\includegraphics[width=1.1in,height=1.25in,clip,keepaspectratio]{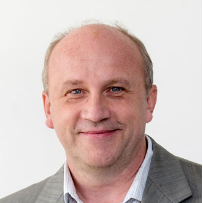}}]{Paul Lukowicz} is a Professor and scientific director of research group Embedded Intelligence at the German Research Center for Artificial Intelligence (DFKI), Germany. His research focuses on cyber-physical systems, pervasive computing and social interactive systems.

\end{IEEEbiography}

\end{document}